\documentclass[9pt,conference]{IEEEtran}
\usepackage{dcase2026}

\usepackage{bm} 

\usepackage{dcase2026,amsmath,graphicx,url,times,booktabs, tabularx}

\usepackage{silence}
\title{EXPLORING FEATURE EXTRACTION TECHNIQUE PARAMETERS FOR ACOUSTIC GUNSHOT CLASSIFICATION}

\name{Sinclair Gurny$^{1}$,
      Ryan Quinn$^{1}$}
\address{$^{1}$Certus Innovations, Albany, NY, USA \;}

\begin{document}

\maketitle

\begin{abstract}
Acoustic gunshot detection is a problem with applications across civilian public safety, military operations, and wildlife conservation, yet the field lacks a rigorous exploration of feature extraction techniques with a focus on generalization to realistic data. The mixed effectiveness of commercial gunshot detection and classification systems indicates an open problem that is not adequately addressed by the current literature. In this paper, we present a systematic investigation of common feature extraction techniques using a dataset of 23,000 gunshot recordings across 85 firearms and 21 calibers. We benchmark three feature extraction techniques with 12 total unique parameter sets using ResNet-18. Our results demonstrate that using the correct feature extraction technique can improve top-1 accuracy by up to 20\%, and utilizing the correct parameters for a given feature extraction technique can improve that value by up to 4.7\%.
\end{abstract}

\begin{IEEEkeywords}
gunshot classification, acoustic classification, feature extraction, deep learning, spectrogram, audio
\end{IEEEkeywords}

\section{Introduction}
\label{sec:intro}

According to the Small Arms Survey, there are more than one billion firearms in circulation worldwide, with the majority in civilian hands \cite{HomeSmallArms2025}. This speaks to the breadth of contexts in which gunshot events can occur, and to the diversity of domains in which automatic acoustic detection systems are needed.
In civilian public safety, rapid detection of firearm discharges can meaningfully reduce emergency response times and support law enforcement operations, though this is complicated by the complex reverberant conditions of indoor and urban environments, where reflected sound waves and background noise can degrade the clarity of the acoustic signature. In military contexts, reliable gunshot classification is essential for situational awareness and threat localization, a task made harder by the wide range of signal-to-noise ratios encountered across environments, from open terrain, the busy environment of a forward-operating base, to dense urban warfare settings \cite{kimMilitaryAudioDataset2024}. In conservation and anti-poaching applications, passive acoustic monitoring represents one of the few scalable means of detecting illegal hunting activity across vast remote areas where continuous human surveillance is logistically infeasible, and where models must perform reliably on sparse, distant recordings captured by unattended sensor hardware \cite{dharFightingPoachingHighPrecision2026}.
Despite the considerable differences between these deployment contexts, they share a common set of technical demands. A robust gunshot classification systems must generalize reliably across firearm types, ammunition varieties, recording conditions, and acoustic environments. It is this requirement for broad generalization, rather than narrow performance in controlled settings, that motivates this paper's exploration of feature extraction techniques.

Research on audio classification with deep learning models has consistently shown that directly processing raw audio underperforms other methods of feature generation, such as spectrograms. These results are consistent across acoustic domains, including but not limited to environmental sound classification \cite{GunshotAudioMuzzle2025} and singing technique classification \cite{yamamotoInvestigatingTimeFrequencyRepresentations2021a}. The precise mechanism by which spectrogram use improves model performance is not formally proven, but credible hypotheses proposed include that the raw audio is excessively information dense \cite{liGunshotRecognitionMethod2022} and that spectrogram generation is a form of dimensionality reduction \cite{razaniReducedComplexityMFCCbased2017}.

ShotSpotter is the most widely deployed commercial gunshot detection system, currently operating across numerous cities in the United States and internationally \cite{aclu_of_oregon_know_2022}. Deployed for nearly 30 years as a networks of acoustic sensors in urban environments meant to detect and triangulate gunshots in real-time for law enforcement, it presents a case study for understanding the practical challenges of gunshot detection and classification. Despite ShotSpotter's history, there is remarkably little publicly available information about the system's internal architecture or independently verified accuracy. A direct field evaluation reported an overall success rate of 80\%, with significant variation across firearm types \cite{FieldEvaluationShotSpotter}.
Furthermore, because short, impulsive sounds are difficult to distinguish in an urban urban environment, ShotSpotter has been specifically criticized for misclassifying fireworks, backfiring vehicles, and construction noises as gunfire \cite{aclu_of_oregon_know_2022}. Therefore, acoustic classification of firearms is far from a solved problem and poses significant systemic challenges when transitioning from laboratory to real-world applications.

In this paper, we present a comprehensive evaluation of the effects of parameter selection in time-frequency representations on deep learning-based acoustic firearm caliber classification. The paper explores the effects of parameter selection on standard short-time Fourier transform spectrograms (STFT), log-mel spectrograms, and mel-frequency cepstral coefficients (MFCCs). We utilize a dataset that comprises five open-access datasets. The resulting dataset is among the largest and most diverse reported in the literature, with over 23,000 total data points across 21 calibers and 85 unique firearms. We perform a total of 12 experiments utilizing ResNet-18 to determine the effect of parameter selection on the performance of the models.

\section{Related Work}
\label{sec:relatedwork}

It has been shown that there is no purely optimal feature extraction technique for acoustic classification; what works best for speech recognition does not necessarily work best for lung sounds \cite{jungEfficientlyClassifyingLung2021}, singing technique identification \cite{yamamotoInvestigatingTimeFrequencyRepresentations2021a}, or bird calls  \cite{carvalhoAutomaticClassificationBird2023}.

Doungpaisan et al. performed a comparative study of CNN architectures and time-frequency representations \cite{doungpaisanDeepSpectrogramLearning2025}. However, parameter sets for each feature extraction method are not explored, and all spectrograms are converted into multichannel RGB images. The dataset is significantly smaller, having only 4 classes with only one firearm per caliber.

Despite some common ground in feature extraction techniques, the exact parameters used differ greatly across the literature.
Raponi, Oligeri, and Ali classified gunshot audio using linear-frequency STFT spectrograms with power dB range scaling, using an FFT window of 65 samples and a hop length of 50\% \cite{raponi_sound_2022}. 
Li et al. classified gunshot audio using mel frequency spectrograms with power dB range scaling, using 40 mel bands, an FFT window of 1024 samples, and a hop length of 50\% \cite{liGunshotRecognitionMethod2022}.
In a subsequent paper, they increased the mel filter banks to 128 and the FFT size to 2048 but kept the 50\% hop length \cite{liFastIdentificationMethod2022}.
Similarly, Shah et al. used 128 mel bands, an FFT window of ~1024 samples, and a hop length of 50\% \cite{shahDecipheringGunTypeHierarchy2025}.
Bajzik et al. used 20 MFCC coefficients derived from 256 mel filter banks, with an FFT length of 2048, and a Hamming window; hop length is not reported \cite{bajzikIndependentChannelResidual2022}.
Most papers do not indicate why the parameters used were chosen, nor whether they explored other options. It is these omission that motivate the current work.

\section{Methods}
\label{sec:data}

\subsection{Datasets}
\label{subsec:dataset}

\begin{table}[t]
\centering
\caption{Data points by caliber count in total dataset}
\label{tab:ammo_counts}
\sisetup{
    reset-text-series = false, 
    text-series-to-math = true, 
    mode=text,
    tight-spacing=true,
    round-mode=places,
    round-precision=2,
    table-format=2.2,
    table-number-alignment=center
}
\begin{tabular}{l r  l r}
    \toprule
    Caliber & Count & Caliber & Count \\
    \midrule
    9x19mm        & 5461 & .357 Magnum        & 343 \\
    .22 LR        & 3364 & 16 Ga              & 330 \\
    .223 Remington & 2038 & 7.62x51mm         & 280 \\
    .45 ACP       & 1577 & 20 Ga              & 275 \\
    5.56x45mm     & 1279 & 7x57mm Mauser      & 140 \\
    12 Ga         & 1209 & 7.62x25 Tokarev    &  94 \\
    .380 ACP      & 1182 & .45-70 Government  &  63 \\
    .40 S\&W      &  986 & .30-06 Springfield &  56 \\
    6.5 Creedmoor &  915 &                 &     \\
    7.62x39mm     &  801 &                 &     \\
    .300 AAC Blackout & 749 &              &     \\
    .38 S\&W      &  702 &                 &     \\
    .38 Special   &  462 &                 &     \\
\end{tabular}
\end{table}

Within the field of acoustic gunshot detection, there are two approaches to collecting datasets: live collection and internet scraping. Live data collection results in higher fidelity data and more descriptive metadata, but the cost of equipment limits the scale. In contrast, datasets scraped from the internet are often larger, but it is harder to verify metadata, and audio quality can suffer due to the unknown video and audio processing steps performed before receiving the data.

There are several notable live-collected datasets available for research. Lilien et al produced two complementary datasets as part of an NIJ-funded project \cite{lilien_development_2018}. The first consisted of controlled test firings of approximately 20 firearms recorded across multiple devices and positions relative to the shooter, while the second comprised audio extracted from YouTube to represent the kind of noisy, real-world recordings that might be encountered in forensics casework. Kabealo et al addressed a well-known gap in earlier datasets by collecting a time-synchronized, multi-orientation dataset at an outdoor firing range using several edge devices positioned around the shooter \cite{kabealoMultifirearmMultiorientationAudio2023}. The Free Firearm Sound Library \cite{bartFreeFirearmSound2014} is an open-source, community-funded sound effects library covering a wide variety of firearms, released under a CC0 license, making it broadly accessible for research use.

In this paper, we utilize a dataset that comprises five open-access datasets. The majority of the dataset, 12,904 samples, were collected by Certus Innovations directly. Around 8,015 samples come from the C3GD Dataset \cite{gurnyCertusCaliberClassification2026} and the remaining 4,889 samples were recently collected and have not been released previously. The remainder of the dataset consists of data collected by Kabealo et al \cite{kabealoMultifirearmMultiorientationAudio2023}, Cadre Forensics \cite{lilien_development_2018}, and The Free Firearm Sound Library \cite{bartFreeFirearmSound2014}.
The full dataset comprises 22,306 recordings across 21 calibers collected with 85 unique firearms. A detailed breakdown of class distributions is provided in \cref{tab:ammo_counts}.

In a full-featured automatic gunshot detection system, the gunshot must be both detected and classified; however, this paper addresses only classification. As evidenced by ShotSpotter's issues with mistakenly identifying other urban sounds, detecting gunshots with high precision and recall in non-laboratory datasets is a difficult problem. A detection dataset would require non-gunshot audio that is relevant for the particular use case. This could consist of construction noises, explosions, and local animals for urban, military, and conservation applications, respectively. Therefore, we chose to focus on classification for its more general nature.

\subsection{Data Augmentation}
\label{subsec:dataaug}

Within the field of image classification, data augmentation is often used to increase the diversity of the dataset, reduce overfitting, and improve generalization. It is often considered to be a part of the standard "bag of tricks" \cite{heBagTricksImage2018}. Despite this, data augmentation is not seen as often within the acoustic gunshot classification field. One of the few examples that shows the utility of this technique is the UrbanSound8K dataset \cite{salamonDeepConvolutionalNeural2017}, which utilizes augmentations such as time stretching, pitch shifting, dynamic range compression, and  background noise. 

This paper includes three data augmentations: time shifting, gaussian noise, and gain adjustment. Time shifting, much like randomized scaling, cropping, and rotation for image classification, is meant to teach invariance to the position of the subject within the data. Gaussian noise is meant to simulate various levels of background noise that may be present in real-world scenarios. It is also possible to include various sounds against which the model should be robust, such as sirens for urban public safety or vehicles in military contexts, but this is not included in this paper for simplicity. Lastly, the gain was randomly adjusted to simulate the volume of gunshots varying due to distance, microphone sensitivity, and occlusion. Notably, pitch shifting and time stretching are not relevant to gunshots, as they distort important frequency and temporal features that are produced by distinct physical phenomena. Data augmentations were implemented using the audiomentations library \cite{GitHubIver56Audiomentations}, and parameters are listed in \cref{tab:tabaug}. The value range for time shifting was chosen based on the configuration of our preprocessing steps, which place the gunshot at 0.1 sec into a 1 sec window.

\begin{table}[t]
\centering
\caption{Data Augmentation Parameters}
\label{tab:tabaug}
\sisetup{
    reset-text-series = false, 
    text-series-to-math = true, 
    mode=text,
    tight-spacing=true,
    round-mode=places,
    round-precision=2,
    table-format=2.2,
    table-number-alignment=center
}
\begin{tabular}{l c c c}
    \toprule
    Augmentation & Min Value & Max Value & Activation Probability \\
    \midrule
    Time Shifting & -0.05 sec & 0.7 sec & 0.90 \\
    Gaussian Noise & 0.001 amp & 0.015 amp & 0.50 \\
    Gain & -6 db & 6 db & 0.75 \\
\end{tabular}
\end{table}

\subsection{Feature Extraction Methods}
\label{subsec:featextmeth}

While there are many ways to generate audio spectrograms; an extensive literature review has found three common options: STFT, log-mel frequency, and MFCC.

The STFT is the foundational operation underlying most acoustic feature representations. By applying a Fourier transform to successive overlapping windows of a raw audio signal, the STFT produces a two-dimensional time-frequency representation, the spectrogram, in which each column captures the instantaneous spectral energy of the signal at a given time. The resulting magnitude spectrogram preserves the full resolution of the linear frequency axis up to the Nyquist frequency, making it sensitive to fine-grained spectral structure across the entire frequency range.

A common perceptually-motivated transformation of the linear spectrogram is the mel spectrogram, in which the frequency axis is warped according to the mel scale, a non linear mapping that approximates the frequency resolution of the human auditory system by applying high resolutions at lower frequencies and coarse resolution at high frequencies \cite{oshaughnessySpeechCommunicationHuman1987}. In practice, this is achieved by passing the STFT magnitude spectrogram through a bank of overlapping triangular filters whose center frequencies are spaced on the mel scale. Applying a logarithmic compression to the filter bank yields the log-mel spectrogram, which additionally accounts for the logarithmic sensitivity of human loudness perception. A review of the literature has shown the log-mel spectrogram to be the dominant input representation in modern audio.

Mel spectrograms have the effect of emphasizing low-frequency features, and have shown excellent performance for environmental and biological sound classification \cite{piczakEnvironmentalSoundClassification2015}. The justification usually given for this enhanced performance is the adaptation of the human ear to natural sounds, so the application of mel filters to gunshot audio, a distinctly non-biological acoustic event, should not be assumed. 

A further derivative of the log-mel spectrogram is MFCCs, which are obtained by applying a discrete cosine transform (DCT) to the log-mel filter bank outputs, producing a compact set of coefficients that capture the slowly-varying spectral envelope of the signal while discarding fine spectral detail and inter-bin correlations \cite{chenPatternRecognitionArtificial1976}. MFCCs were the standard representation in classical speech and audio recognition systems and remain widely used due to their low dimensionality and robustness to certain forms of acoustic variability. MFCCs are alternatively said to be robust to noise \cite{RepresentingAudioData} and susceptible to it \cite{tyagiDesensitizingMelcepstrumSpurious2005}, bringing into question their applicability to the analysis of field recordings. 

Additionally, the justification for using MFCCs as a feature extraction method is often elided or given simply as “they are popular.” Khan writes that the “main principle behind MFCC is to condense essential information into a concise set of coefficients, inspired by the human ear’s auditory perception” \cite{khanIndoorGunshotDetection2023}. No reference for this claim is provided, and other papers have noted that human perception may not be  well matched to gunshots \cite{sigmundEfficientFeatureSet2021}.

\subsection{Feature Extraction Parameters}
\label{subsec:featextpars}

The choice of FFT window length has significant implications for the resulting spectrogram, as it controls the trade-off between time and frequency resolution. Despite this, we have found few papers that investigate the effects of this parameter on model performance, notably \cite{yamamotoInvestigatingTimeFrequencyRepresentations2021a}. 
To avoid spectral leakage, a windowing function must be chosen; a review of the literature showed a split between Hann \cite{raponi_sound_2022} and Hamming \cite{liFastIdentificationMethod2022} \cite{bajzikIndependentChannelResidual2022}  \cite{doungpaisanDeepSpectrogramLearning2025} windows. 

The other notable parameter, hop length, must be set to ensure that data are weighted equally and to avoid data aliasing \cite{ChoiceHopSize}. Papers often elide their choice of hop length \cite{bajzikIndependentChannelResidual2022}, or use a 50\% window overlap without explanation \cite{raponi_sound_2022} \cite{shahDecipheringGunTypeHierarchy2025}. For a Hann or Hamming window, a 75\% overlap is mathematically ideal \cite{ChoiceHopSize}, so this divergence bears investigating.

In addition to the parameters needed to apply an STFT, a mel spectrogram also needs to select a number of mel filter banks. A literature review did not return definitive suggestions, with values between 40 and 256 returning reasonable results \cite{liGunshotRecognitionMethod2022}, \cite{bajzikIndependentChannelResidual2022}, and the TorchAudio library using a default of 128.
While they do not give a justification for using mel frequency spectrograms over STFTs, Li et al. \cite{liGunshotRecognitionMethod2022} \cite{liFastIdentificationMethod2022} show excellent results from their use. 
Given the lack of clarity on the ideal parameters for mel spectrograms contrasted with the plethora of published results, this knowledge gap bears exploration. 

The justification for parameter set(s) for MFCCs are often not reported, particularly the number of cepstral coefficients used. Numbers vary between 13 \cite{nesarMachineLearningAnalysis2024a}, 20-40 \cite{jungEfficientlyClassifyingLung2021} and 40 \cite{singhMeasurementsAnalysisClassification2022}. Overall, the literature points to the conclusion that the optimal number of cepstral coefficients is an open question.

For speech enhancement, Razani et al. found that MFCC features slightly out-performed STFT features and led to shorter training times and reduced complexity \cite{razaniReducedComplexityMFCCbased2017}.
MFCCs have been compared against STFTs outside the domain of speech analysis with mixed results. Jung et al. compared them for lung pathology classification without explanation and found that MFCCs significantly underperformed \cite{jungEfficientlyClassifyingLung2021}. For industrial sounds, the performance of STFT vs. MFCC features has been shown to be inconsistent \cite{bajzikIndependentChannelResidual2022}. Bajzik et al. investigated MFCCs vs. log-mel spectrograms and found a significant advantage for the latter that increased with sample rate \cite{bajzikIndependentChannelResidual2022}.
Given the combined popularity and dearth of definitive results, the efficacy of MFCC features for gunshot classification must be revisited. 

\section{Experiments}
\label{sec:experiments}

In this work, we evaluate several possible configurations of these feature representations as inputs to our classification model. Different spectrogram parameters alter the balance of frequency and temporal resolution, and the trade-offs between the two for gunshot classification have not been previously studied. We utilize at least three parameter sets for each spectrogram type. The experiments were performed on a workstation with 64 GB of RAM, an Intel i9-12900K, and a NVIDIA GeForce GTX 4060. Training code and training data is available\footnote{https://github.com/Stonewall-Defense/certus-dcase-2026-training-code}.

A detailed enumeration of the parameters used is provided in \cref{tab:tabexp}. In addition to parameter sets that favor time vs. frequency resolution, we include a parameter set called "Hann Ideal," which utilizes a hop length for mathematically optimal side lobe suppression, detailed in \cref{subsec:featextmeth}. MFCCs do not have such a dichotomy, but the number of MFCC values used varies largely in research. Therefore, we selected three values which can be loosely categorized as small, medium, and large with the exact values being 20, 30, and 40. The parameters were not drawn from a particular literature source; they are an attempt to explore as much of the previously-reported configuration space as was reasonable.

The model architecture used was ResNet-18, specifically the implementation from \href{https://timm.fast.ai/}{TIMM}. The inputs were modified for single-channel data, and the classifier head was modified for 21 output classes. The model was trained for 100 epochs with a learning rate of 0.0005 and a batch size of 32, which took how approximately one hour on our machine. All code was written for \href{https://pytorch.org/}{PyTorch} compatibility, using the \href{https://lightning.ai/docs/pytorch/stable/}{Lightning} library for training. Each experiment was run five times with different random seeds; each experiment used the same random seeds as the others in the same order.

\begin{table}[t]
\centering
\caption{Experiment Parameter Sets}
\label{tab:tabexp}
\sisetup{
    reset-text-series = false, 
    text-series-to-math = true, 
    mode=text,
    tight-spacing=true,
    round-mode=places,
    round-precision=2,
    table-format=2.2,
    table-number-alignment=center
}
\begin{tabular}{c c c c c c c}
    \toprule
    Type & Focus & FFTs & Hop Len & Scale & Mels & CCs \\
    \midrule
    Log-Mel & Balanced & 1024 & 512 & log & 128 & - \\
    Log-Mel & Freq & 2048 & 1024 & log & 256 & - \\
    Log-Mel & Hann Ideal & 1024 & 256 & log & 128 & - \\
    Log-Mel & Time & 512 & 256 & log & 64 & - \\
    \midrule
    MFCC & Small & 1024 & 512 & log & 128 & 20 \\
    MFCC & Medium & 1024 & 512 & log & 128 & 30 \\
    MFCC & Large & 1024 & 512 & log & 128 & 40 \\
    \midrule
    STFT & Balanced & 1024 & 512 & log & - & - \\
    STFT & Freq & 2048 & 1024 & log & - & - \\
    STFT & Hann Ideal & 1024 & 256 & log & - & - \\
    STFT & Linear & 1024 & 512 & linear & - & - \\
    STFT & Time & 512 & 256 & log & - & - \\
\end{tabular}
\end{table}

\section{Results}
\label{sec:results}

\begin{figure}[t]
  \centering
  \centerline{\includegraphics[width=\columnwidth]{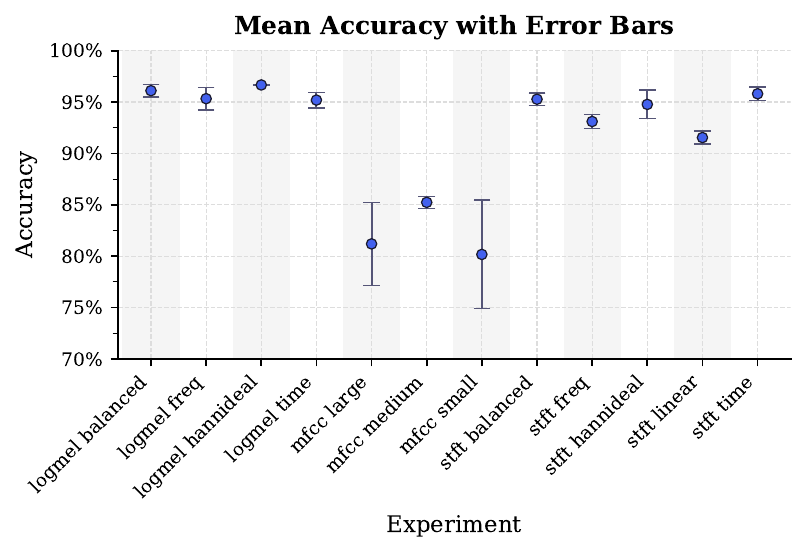}}
  \caption{Results of each feature extraction experiment}
  \label{fig:results}
\end{figure}

\cref{tab:full_results} and \cref{fig:results} report the performance of all three methods across parameters configurations. Overall, log-mel spectrograms achieved the best results with a maximum accuracy of 96.66\%, followed by STFT spectrograms at 95.81\%, and MFCCs at 85.24\%. The averages across each method, \cref{tab:method_results}, show the same ordering.

MFCC-based features significantly underperformed other options, both in terms of accuracy and consistency. The relative, though not statistically significant, out-performance of 30 cepstral coefficients may indicate a nonlinear relationship between spectral information retention and feature simplification; more research is needed to confirm this pattern.

Mesaros et al note that MFCCs provide "both compression and decorrelation of the information in the signal spectrum" \cite{mesarosSoundEventDetection2021}. They note that decorrelation was beneficial to GMMs and other early ML models, but has no effect on CNNs because they can learn from correlated data. Our results show that whatever representational benefits MFCCs may have provided previously, they do not transfer well to ResNets for gunshot classification.

The consistency of log-mel spectrogram results between the time, balanced, and frequency parameter sets is consistent with previous work by other groups showing promising results from a wide range of mel band choices. The only significant result among log-mel spectrograms was the “Hann Ideal” parameter set, which also shows the least variability between experiments. These results show that the often-overlooked hop length parameter is an important component of model results.

For STFT feature tests, linear-amplitude spectrograms performed statistically significantly worse than other feature sets. We found no examples of linear-amplitude spectrograms in the literature; this demonstrated under-performance may indicate why that is. Since the spectrograms are scaled to a range of [0, 1] before processing, the higher initial dynamic range of linear-amplitude spectrograms may have obscured important differences in the data.

Within dB-scaled STFT experiments, there is a direct correlation between time resolution and accuracy, with the "STFT-Freq" parameter set performing statistically significantly worse than other experiments. Gunshot audio is known to have distinctive time-domain characteristics that vary by muzzle velocity, bullet diameter, and mic orientation \cite{lilien_development_2018} \cite{kabealoMultifirearmMultiorientationAudio2023} \cite{nesarMachineLearningAnalysis2024a}. Our data show that time-dependent gunshot audio features may contain more information than frequency-based ones, which is consistent with the high performance of the log-mel results that condense frequency information into only 1/4 of the bins of the STFT spectrograms.

Mel spectrograms and STFT results were generally consistent, with the balanced and Hann Ideal STFT parameter sets statistically indistinguishable from the mel results. These results imply that while similar amounts of information are contained in the low- and high-frequency portions of the gunshot spectrogram, log-mel spectrograms show more consistent performance across a range of parameter sets. More research is needed to confirm these conclusions. 

\begin{table}[t]
\centering
\caption{Results of each feature extraction experiment}
\label{tab:full_results}
\sisetup{
    reset-text-series = false, 
    text-series-to-math = true, 
    mode=text,
    tight-spacing=true,
    round-mode=places,
    round-precision=4,
    table-format=1.4,
    table-number-alignment=center
}
\begin{tabular}{l l S S[round-precision=4,table-format=1.4] S S}
    \toprule
    Method & {Focus} & {Mean Acc} & {Std Dev} & {Min} & {Max} \\
    \midrule
    LogMel & Balance   & 0.961071 & 0.006173 & 0.956250 & 0.971875 \\
    LogMel & Freq       & 0.953304 & 0.010961 & 0.933929 & 0.959821 \\
    LogMel & Ideal & \bfseries 0.966635 & 0.000370 & 0.966071 & 0.966964 \\
    LogMel & Time       & 0.951931 & 0.007706 & 0.940766 & 0.958482 \\
    \midrule
    MFCC & Small  & 0.801715 & 0.052881 & 0.726316 & 0.865907 \\
    MFCC & Medium & \bfseries 0.852401 & 0.005781 & 0.847768 & 0.861748 \\
    MFCC & Large  & 0.811983 & 0.040533 & 0.743139 & 0.842105 \\
    \midrule
    STFT & Balance   & 0.952768 & 0.006001 & 0.946429 & 0.959375 \\
    STFT & Freq       & 0.931071 & 0.006767 & 0.925446 & 0.942411 \\
    STFT & Ideal & 0.947730 & 0.013955 & 0.927796 & 0.964732 \\
    STFT & Linear     & 0.915446 & 0.006334 & 0.909821 & 0.922768 \\
    STFT & Time       & \bfseries 0.958064 & 0.006697 & 0.950893 & 0.965320 \\
\end{tabular}
\end{table}

\begin{table}[t]
\centering
\caption{Results reduced to each feature extraction method}
\label{tab:method_results}
\sisetup{
    reset-text-series = false, 
    text-series-to-math = true, 
    mode=text,
    tight-spacing=true,
    round-mode=places,
    round-precision=4,
    table-format=1.4,
    table-number-alignment=center
}
\begin{tabular}{l S S[round-precision=4,table-format=1.4,] S S}
    \toprule
    Method & {Mean Acc} & {Std Dev} & {Min} & {Max} \\
    \midrule
    Log Mel & \bfseries 0.958235 & 0.009132 & 0.933929 & 0.971875 \\
    MFCC    & 0.822033 & 0.042318 & 0.726316 & 0.865907 \\
    STFT    & 0.941016 & 0.017770 & 0.909821 & 0.965320 \\
\end{tabular}
\end{table}

\section{Conclusion}
\label{sec:conclusion}

This paper systematically investigated the most common feature extraction techniques and their parameters used in acoustic gunshot classification, with a focus on generalization to real-world situations. The results show that choosing the correct parameters for a particular feature extraction method can often provide a statistically significant improvement in model performance. Log-mel spectrograms outperformed STFTs slightly and MFCCs significantly. Notably, MFCCs had the largest range of results; it has been stated that MFCCs are not as robust to noise as other feature extraction techniques, which may explain the results. There are no clear trends between parameter sets and model performance, suggesting that task-specific tuning remains essential.

Overall, these findings imply that no single feature extraction approach dominates universally; finding the best technique and parameters should be a core step in model development.

While this paper focused on the core set of feature extraction techniques used in acoustic gunshot classification, a natural extension is broadening this investigation to include additional methods showing promise in the wider acoustic classification literature. Furthermore, it is also possible to pass multi-channel input into models that consists of different spectrograms or parameters sets as each channel.

The supersonic shockwave produced by the bullet provides information potentially relevant to detection, classification, and localization simultaneously, but is often missing from open datasets, including ours. Data collection efforts should prioritize the supersonic shockwave with highly detailed metadata and annotation, as it remains underexplored using modern techniques.

\section{Acknowledgment}
\label{sec:ack}

This work was funded in part by Air Force Research Laboratory contract FA8750-24-C-B082.
All authors reviewed the manuscript. The article authors have declared no conflicts of interest.

\clearpage

\bibliographystyle{IEEEtran}
\bibliography{refs}

\end{document}